\documentclass[prl,twocolumn,showpacs,preprintnumbers,amsmath,amssymb,citeautoscript]{revtex4-1}

\usepackage{graphicx}% Include figure files
\usepackage{dcolumn}% Align table columns on decimal point
\usepackage{bm}% bold math

\begin{document}

\title{Thermal conductivity of quantum magnetic monopoles in the frustrated pyrochlore Yb$_2$Ti$_2$O$_7$}

\author{Y. Tokiwa$^{1,2}$}
\author{T. Yamashita$^1$}
\author{M. Udagawa$^3$}
\author{S. Kittaka$^4$}
\author{T. Sakakibara$^4$}
\author{D. Terazawa$^1$}
\author{Y. Shimoyama$^1$}
\author{T. Terashima$^2$}
\author{Y. Yasui$^5$}
\author{T. Shibauchi$^6$}
\author{Y. Matsuda$^1$}

\affiliation{$^1$Department of Physics, Kyoto University, Kyoto 606-8502, Japan}
\affiliation{$^2$Research Center for Low Temperature and Materials Science, Kyoto University, Kyoto 606-8501, Japan}
\affiliation{$^3$Department of Applied Physics, University of Tokyo, Bunkyo, Tokyo 113-8656, Japan}
\affiliation{$^4$Institute for Solid State Physics, University of Tokyo, Kashiwa 277-8581, Japan}
\affiliation{$^5$Department of Physics, School of Science and Technology, Meiji University, Higashi-mita, Tama-ku, Kawasaki 214-8571, Japan}
\affiliation{$^6$Department of Advanced Materials Science, University of Tokyo, Chiba 277-8561, Japan}

\begin{abstract}

We report low-temperature thermal conductivity $\kappa$ of pyrochlore Yb$_2$Ti$_2$O$_7$, which contains frustrated spin-ice correlations with significant quantum fluctuations.  In  the disordered spin-liquid regime, $\kappa(H)$ exhibits a nonmonotonic magnetic field dependence, which  is well explained by the strong spin-phonon scattering and quantum monopole excitations.  We show that the excitation energy of quantum monopoles is  strongly suppressed from that of dispersionless classical monopoles.  Moreover, in stark contrast to the diffusive classical monopoles, the quantum monopoles have a very long mean free path. We infer that the quantum monopole is a novel heavy particle, presumably boson, which is highly mobile in a three-dimensional spin liquid.

\end{abstract}
\maketitle

Rare-earth pyrochlore oxides exhibit various exotic magnetic properties owing to their strong geometrical frustration experienced by coupled magnetic moments on the tetrahedral lattice (Fig.\,1(a))~\cite{Bramwell-Science01}. The most explored materials are Ho$_2$Ti$_2$O$_7$ and Dy$_2$Ti$_2$O$_7$, in which the magnetic moments can be regarded as classical spins with a strong easy-axis (Ising) anisotropy~\cite{Ramirez-Nature99,Bramwell-Science01}. The frustration of these moments results in a remarkable spin ice with macroscopically degenerate ground states,  in which each tetrahedron has the ``two spins in, two spins out (2-in-2-out)" configuration.  This  spin structure is characterized by dipolar spin correlations with a power-law decay, which is observable as the unusual pinch-point shape of spin structure factor by neutron scattering~\cite{Bramwell-PRL01,Fennell-Science09}. One of the most remarkable features of the spin-ice state is that it hosts emergent magnetic monopole excitations; the first excitation is 3-in-1-out configuration~\cite{Castelnovo-Nature08,Morris-Science09}. This  produces a bound pair of north and south poles, which can be fractionalized into two free magnetic monopoles. This classical monopole excitations are gapped and dispersionless (Fig.\,1(b)). Therefore the propagation of monopoles occurs only diffusively and the monopole population decays exponentially at temperatures well below the gap.  Of particular interest is how the spin-ice ground state is altered by the quantum fluctuations, which may lift  the degeneracy of the spin-ice manifold, leading to a new ground state such as quantum spin-ice state~\cite{Shannon-PRL12,Benton-PRB12,Gingras-RPP14,Savary-PRL12,Harmele-PRB04}. To clarify this issue, uncovering newly emergent elementary excitations in the presence of quantum fluctuations is crucially important.  Although exotic excitations such as gapless photon-like mode have been proposed theoretically, the nature of excitations are poorly explored.  

Among the magnetic pyrochlore materials, Yb$_2$Ti$_2$O$_7$, Er$_2$Ti$_2$O$_7$, Pr$_2$Sn$_2$O$_7$  and possibly Tb$_2$Ti$_2$O$_7$ host  strong transverse quantum fluctuations of magnetic dipoles owing to speudospin-1/2 of magnetic rare earth elements~\cite{Ross-PRX11,Applegate-PRL12,Hirschberger-15,Gardner-RMP10}.  In particular, Yb$_2$Ti$_2$O$_7$ with well separated crystalline electric-field excited levels is a good model system to study the influence of the quantum effects on monopole excitations~\cite{Hodges-JPhys01}.  Yb$_2$Ti$_2$O$_7$ undergoes a weakly first-order ferromagnetic phase transition at $T_{\rm C}\sim 0.2$\,K~\cite{chang-NComm12,Yasui-JPSJ03,Ross-PRB11,Hodges-PRL02}. It is widely believed that quantum fluctuations melt the spin-ice state into a spin-liquid state. Therefore, the pinch point structure observed by neutron scattering above $T_{\rm C}$ indicates the presence of a spin-liquid phase with spin-ice correlations~\cite{chang-NComm12}.  The full set of Hamiltonian parameters is determined by inelastic neutron scattering experiments, providing a prototypical system described by an effective pseudospin-1/2 quantum spin-ice model~\cite{Ross-PRX11}. The Hamiltonian consists of  three main interactions, $J_{\parallel}$, $J_{\perp}$ and $J_{z\pm}$. Here $J_{\parallel}(=2$\,K) is the Ising component of the nearest neighbor interaction, $J_{\perp}(=0.58$\,K) is the XY-component and  $J_{z\pm}(=1.7$\,K) is the off-diagonal component. Finite $J_{\perp}$ and $J_{z\pm}$ produce quantum fluctuations (Fig.\,1(c)).  

Here, to study the elementary excitations in the spin liquid state of Yb$_2$Ti$_2$O$_7$,  we measured the thermal conductivity, which is a powerful probe for low energy excitations  at low temperatures, providing a sensitive measurement of a flow of entropy conducted by magnetic excitations and phonons.    The thermal conductivity has been reported in the classical spin-ice state of Dy$_2$Ti$_2$O$_7$ recently. However, the interpretation of the thermal conductivity of Dy$_2$Ti$_2$O$_7$ appears to be complicated owing to the strongly suppressed phonon thermal conductivity by unknown additional scatterings (see Supplemental Material~\cite{supple}). In fact, suggested heat transport by classical monopole is at odds with the diffusive motion of the dispersionless classical monopoles.  
%and a substantial heat conduction due to classical monopoles has been suggested.  In fact, the suggested long mean free path of classical monopoles obviously contradicts the diffusive motion of the dispersionless classical monopoles.  
We show that the thermal conductivity of Yb$_2$Ti$_2$O$_7$  is rather simple: the phonon term shows a $B/T$ scaling and the monopole contribution vanishes below $T_{\rm C}$ as expected.  Our analysis shows the evidence of  the substantial heat transport by quantum magnetic monoples, whose excitation energy is significantly suppressed from that of classical monopoles.   The quantum magnetic monopoles are highly mobile due to quantum fluctuations, in stark contrast to the localized and diffusive nature of classical monopoles. 

\begin{figure}[t]
\includegraphics[width=\linewidth,keepaspectratio]{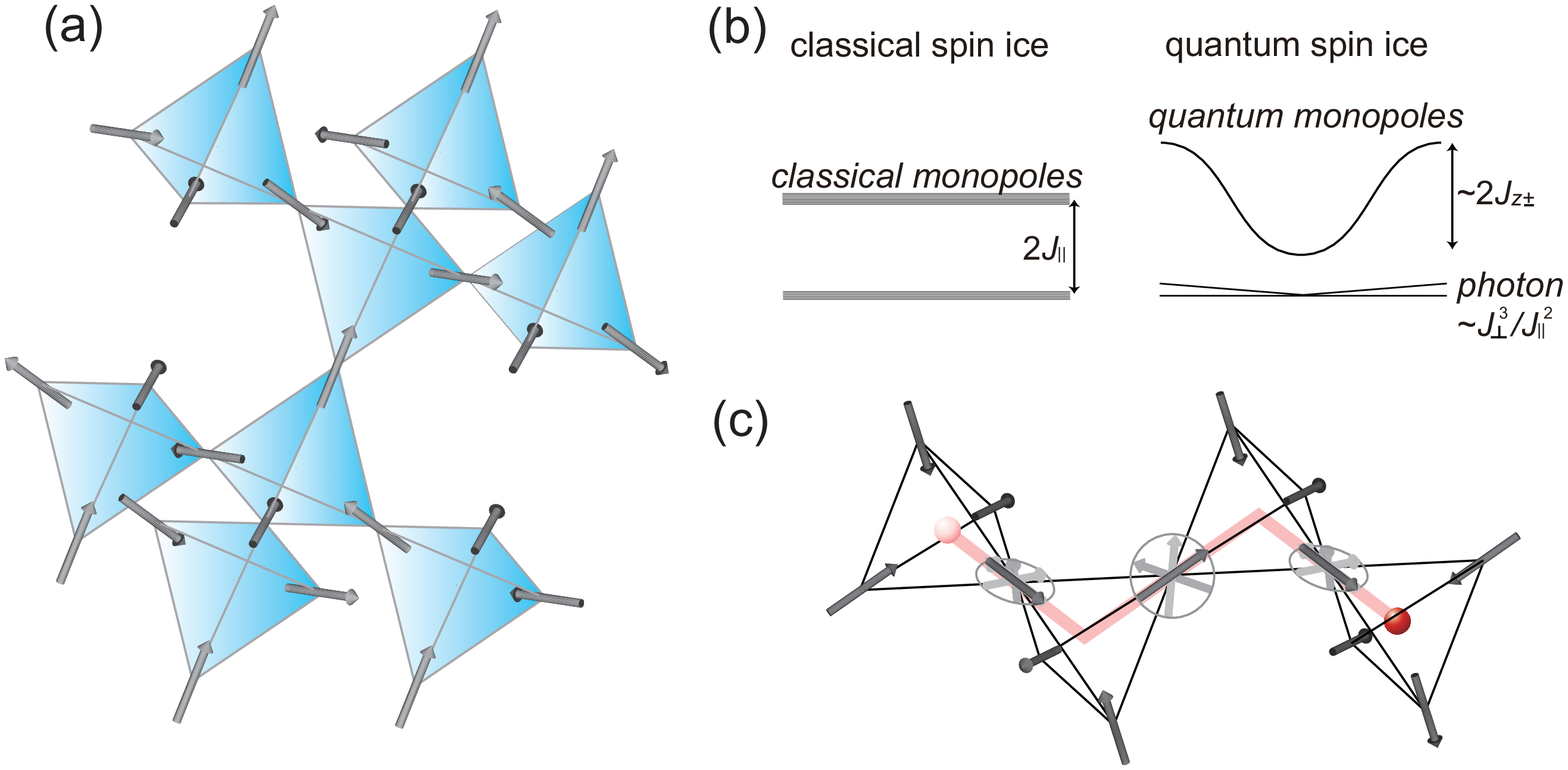}
\caption{(color online). (a) Spin-ice structure on frustrated pyrochlore lattice. (b) Magnetic monopole excitations in classical and quantum spin ice. In classical spin ice the gap energy is twice the Ising interaction of magnetic moments 2$J_{\parallel}$. In the quantum spin ice the off-diagonal interaction $J_{z\pm}$ gives rise to a dispersive monopole excitation. The photon excitations based on the $XY$-component $J_{\perp}$ lift the ice degeneracy in the ground state. (c) Collective motion of quantum magnetic monopoles.}
\end{figure}

High quality single crystals of Yb$_2$Ti$_2$O$_7$ were grown by the floating zone method. Thermal conductivity was measured along [1,-1,0] direction by the standard steady-state method in a dilution refrigerator. Magnetic field was applied along [1,1,1] and [0,0,1], perpendicular to the heat current. Specific heat was determined by the quasi-adiabatic heat pulse method in a dilution refrigerator.

\begin{figure}[t]
\includegraphics[width=\linewidth,keepaspectratio]{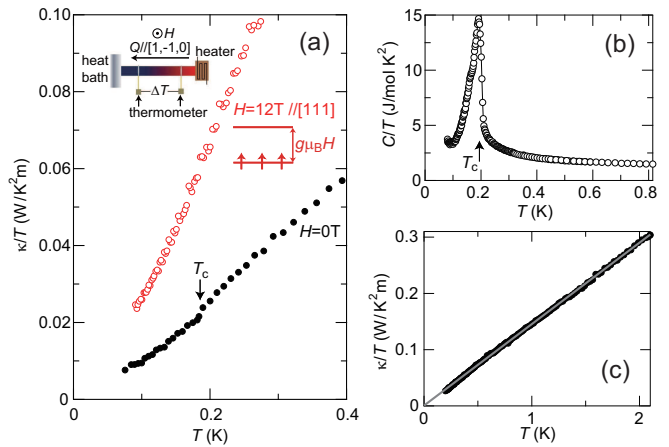}
\caption{(color online). (a) $\kappa/T$ at zero and $\mu_0H=12$\,T applied along [1,1,1] direction is plotted against temperature. The heat current is applied along [1,-1,0].  At $T_{\rm C}$, $\kappa(0)/T$ exhibits a jump, indicated by an arrow. Inset illustrates the measurement configuration of the thermal conductivity. (b) Specific heat divided by temperature $C/T$ at zero field. (c) $\kappa/T$ at zero field in the spin liquid state above 0.2\,K. Grey line is a fit to a $T$-linear dependence $\kappa$/$T=AT$ with $A=0.15$\,W/K$^2$m.
}
\end{figure}

Figure\,2(a) shows the temperature dependence of thermal conductivity divided by temperature $\kappa/T$ in zero field and at $\mu_0H=12$\,T measured on a single crystal of Yb$_2$Ti$_2$O$_7$. Distinct jump in $\kappa/T$ at zero field is observed at $T_{\rm C}$. As shown in Fig.\,2(b), the specific heat $C$ of the single crystal taken from the same batch shows a sharp and large jump at $T_{\rm C}$~\cite{chang-NComm12}. We note that in the previous studies, such a sharp single jump in $C/T$ had been reported only in the powered samples~\cite{Yaouanc-PRB11,Ross-PRB11}, demonstrating the high quality of the present crystal.  We also note that the pinch point features in neutron scattering has been clearly observed in the single crystal which shows a similar specific heat jump~\cite{chang-NComm12}. As shown in Fig.\,2(c), zero-field $\kappa/T$ above $T_{\rm C}$ follows a $T$-linear dependence with negligibly small intercept at $T= 0$\,K. The absence of residual $\kappa /T\mid_{T\to 0\,{\rm K}}$ in the spin-liquid state with spin-ice correlations will be discussed later.

As clearly seen in Fig.\,2(a), magnetic field strongly enhances the thermal conductivity.  Figures\,3(a) and 3(b) and their insets show the field dependence of $\kappa(H)/T$ for different field directions.  %These field dependencies of $\kappa(H)/T$ are distinctly different from those reported in Dy$_2$Ti$_2$O$_7$~\cite{Kolland-PRB12}. 
 As illustrated in the inset of Fig.\,3(c), there are three characteristic regimes; low-field regime (i) where $\kappa(H)/T$ decreases with $H$, intermediate field regime (ii) where $\kappa(H)/T$ increases, and high-field regime where $\kappa(H)/T$ exhibits a saturation.  

In the present system, heat is transferred by phonons and magnetic excitations: $\kappa=\kappa_p+\kappa_m$.  We point out that the field dependence of $\kappa(H)/T$ in the (ii) and (iii) regimes are dominated by the phonon contribution $\kappa_p$ determined by spin-phonon scattering, which contains elastic and inelastic processes.  The elastic scattering (determined by $J_\parallel$) is enhanced with increasing disorder of spins and thus this scattering process should be monotonically suppressed by the alignment of spins with increasing magnetic field. A recent calculations of magnetoresistance  in a fluctuating spin-ice state indicates that the electron-spin elastic scattering rate decreases with increasing magnetization \cite{Udagawa}, which supports this trend. The inelastic  scattering is directly related to the quantum dynamics of spin. In this inelastic scattering, the leading spin-flip process accompanies a hopping of a monopole to the neighboring tetrahedron (which is related to $J_{\perp}$), because this process requires much lower energies than creation or annihilation of monopoles. This scattering is suppressed with field by the formation of Zeeman gap.  Therefore an external magnetic field suppresses both elastic and inelastic scatterings, leading to the enhancement of the phonon thermal conductivity $\kappa_p$.   

In the regime (iii), the Zeeman splitting energy $g\mu_BH$ well exceeds both of the magnetic interactions and thermal energy, $g\mu_BH\gg J_{\parallel}, J_{\perp}, J_ {z\pm}$ and $k_BT$.    In this situation, where all spins are fully polarized and the magnetic (spin-wave) excitations are gapped with a gap $g\mu_BH$, thermal conductivity is almost entirely dominated by the pure phonon contribution without spin scattering because of the following reasons. First, elastic spin-phonon scattering is absent due to the perfect alignment of spins.  Second, inelastic scattering is also absent due to the formation of the large Zeeman gap.  Third, spins do not carry the heat due to the Zeeman gap.  Since purely phononic thermal conductivity is insensitive to magnetic field, $\kappa(H)/T$ in the regime (iii) is nearly independent of $H$.   In the regime (ii),  the phonon mean free path is significantly reduced by the spin-phonon scattering due to the spins thermally excited across the Zeeman gap.  In fact, as shown in Fig.\,3(c) which plots $\kappa/T$ as a function of $\mu_BH/k_BT$, all data collapse into a single curve except for the low $\mu_BH/k_BT$ regime.  The fact that data for both field directions stabilizing different spin configurations (3-in-1-out for $\bm{H}\parallel [1,1,1]$ and 2-in-2-out for $\bm{H}\parallel [0,0,1]$) follow the same curve implies that the elastic spin-phonon scattering dominates over the inelastic scattering in this regime.  It is intriguing that the $H/T$ scaling curve appears to follow the Brillouin function (the dashed line in Fig.\,3(c)).  Here we fitted the data with $g$-factor of 0.79, which is comparable to Land\'{e} $g$-factor (8/7) of Yb.  This coincidence with the Brillouin function calls for further theoretical investigations. 

\begin{figure}[t]
\includegraphics[width=\linewidth,keepaspectratio]{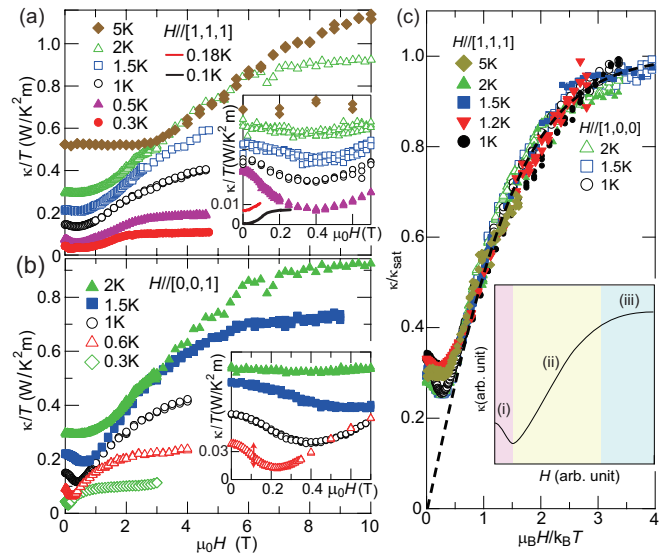}
\caption{(color online). (a) Field dependence of $\kappa/T$ of Yb$_2$Ti$_2$O$_7$ for {\boldmath $H$} $\parallel$ [1,1,1]  with the heat current along [1,-1,0]. The inset shows $\kappa(H)/T$ at low field. Solid red and black lines are $\kappa/T$($H$) at 0.18 and 1\,K, respectively. Data are shifted vertically for clarity.  (b) The same plot for  {\boldmath $H$} $\parallel$ [0,0,1].   Double-headed red arrow in the inset of (a) indicates the initial reduction of $\kappa(H)/T$ for {\boldmath $H$} $\parallel$[0,0,1] at $T$=0.6\,K, which is estimated to be 0.03 \,W/K$^2$m, giving a lower-bound estimate of the monopole contribution.  (c) Normalized thermal conductivity $\kappa/\kappa_{\rm sat}$ plotted against $\mu_{\rm B}H/k_{\rm B}T$, where $\kappa_{\rm sat}$ is the saturated thermal conductivity at high fields.  $\kappa_{\rm sat}$ at high temperatures is determined so as to fit the scaling curve. The dashed line represents the Brillouin function with spin=1/2, assuming $g$=0.79.   The inset illustrates the typical behavior of $\kappa(H)/T$.  There are three characteristic field regimes, (i),(ii) and (iii) indicated by different colors.}
\end{figure}

A particularly important information for the elementary excitations is provided by $\kappa(H)/T$ in the regime (i),  where $\kappa(H)/T$ decreases with $H$ (the insets of Figs.\,3(a) and (b)) and exhibits striking deviations from the $H/T$-scaling curve (Fig.\,3(c)).  This low-field behavior of $\kappa(H)/T$ is most likely due to the monopole contribution $\kappa_m$ because of the following reasons.  First, the initial reduction with $H$ cannot be explained by spin-phonon scattering, which always increases $\kappa(H)$ with $H$ as discussed above.  Second, the deviations in regime (i) appear below  $T^*\sim 4$\,K, where the pinch point features in neutron scattering appears~\cite{chang-NComm12}.  In addition, $T^*$ is close to the temperature $2J_{\parallel}/k_B$, above which monopole excitation disappears.  Third, as shown in the inset of Fig.~3(a) (see Supplemental Material~\cite{supple} for more detail), the initial reduction of $\kappa(H)/T$ disappears below $T_{\rm C}$, which is consistent with the monopole scenario because ferromagnetic ordering prevents the monopole formation. 

\begin{figure}[t]
\includegraphics[width=\linewidth,keepaspectratio]{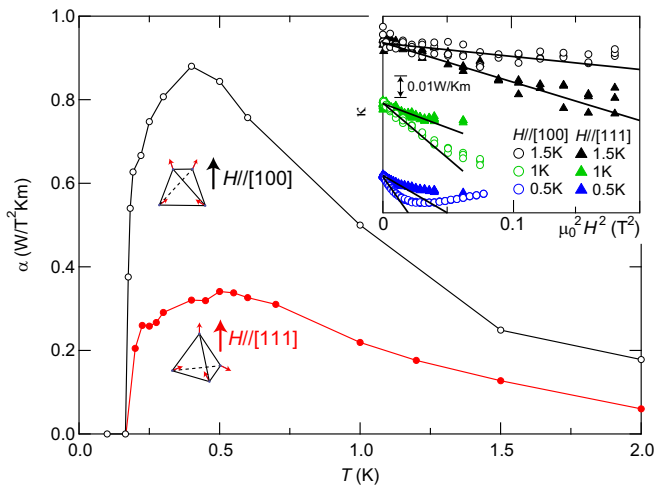}
\caption{(color online). Temperature dependence of the initial slope of $\kappa(H)$ determined by fitting $\kappa(H)=\kappa(0)-\alpha H^2$, for {\boldmath $H$}$\parallel$ [0,0,1] and [1,1,1].   Inset shows $\kappa(H)$  plotted as a function of $H^2$ at very low field. }
\end{figure}

The decrease of $\kappa(H)/T$ with $H$ implies that the number of monopoles is reduced with $H$ at low fields. This reduction is expected even in the dispersionless classical monopoles with gap $2J_{\parallel}$ (see Supplemental Material~\cite{supple}).  However, in the classical case, the number of monopoles will decay exponentially with decreasing temperature below $T^*\sim 2J_{\parallel}/k_B$.   Therefore the observed quite substantial reduction of $\kappa(H)/T$ even at low temperatures well below $T^*$  is totally inconsistent with the classical monopoles.  The results provide strong evidence that the monopole excitation gap is dramatically suppressed from the classical monopole, suggesting the emergence of dispersive quantum magnetic monopoles illustrated in Figs.\,1(c) and (d).  We also note that the substantial reduction of monopole density by the low field will result in a reduction of the inelastic spin-phonon scattering process related to the monopole hopping discussed above, which further emphasizes the significant role of the quantum monopoles themselves as a heat conducting carrier at low fields.

As shown in the inset of Fig.\,4, $\kappa(H)$ decreases as $\kappa(H)=\kappa(0)-\alpha H^2$ $(\alpha>0)$ at very low fields. As the thermal conduction by magnetic excitations  is determined by the number of low-energy itinerant quasiparticles, this $\alpha$ is a measure of the suppression rate of magnetic monopoles at low fields.  Figure\,4 depicts the temperature dependence of $\alpha$ for $\bm{H}\parallel$ [0,0,1] and [1,1,1].  As the temperature is lowered, $\alpha$ first increases, decreases after showing maximum at $T_{\rm max}=0.3$-0.5\,K and suddenly vanishes at $T_{\rm C}$.   The difference in the magnitude of $\alpha$ in the two field directions may be related to the expected difference in the density of 3-in-1-out configuration at high fields, but the trends of $\alpha(T)$ are similar in both cases. The enhancement of $\alpha$ with decreasing $T$ below $T^*$ can be accounted for by the reduction of thermal smearing of the monopole excitations.  The reduction of $\alpha$ at lower temperatures is expected when the thermal energy scale $k_BT_{\rm max}$ becomes comparable to a fraction of the monopole excitation gap (see Supplemental Material~\cite{supple}). In addition, possible ferromagnetic fluctuations, because of the weakly first-order nature of the transition at $T_{\rm C}$, would also suppress $\alpha$ near $T_{\rm C}$. In any case, the observed low energy scale of $k_BT_{\rm max}$ indicates that the excitation gap is strongly reduced than the classical monopole case ($2J_{\parallel} =4$\,K).  The present results lead us to conclude that the thermally excited quantum monopoles carry substantial portion of the heat particularly in the regime (i).  This is reinforced by the fact that $\kappa/T$ at zero field shows a distinct decrease below $T_{\rm C}$ (Fig.\,2), where the phonon contribution $\kappa_p$ is expected to be enhanced owing to the ferromagnetic spin alignment.  

Next we demonstrate that  quantum monopoles are highly mobile in the crystal lattice. Assuming the kinetic approximation, the monopole contribution to the thermal conductivity $\kappa_m$ is written as $\kappa_m=C_mv\ell/3$, where $C_m$ is the monopole contribution in the specific heat,  $v$ is the velocity and $\ell$ is the mean free path of the monopoles.  We estimate $\ell$ at 0.6\,K simply by assuming that the amount of  initial reduction of $\kappa(H)/T$ shown red double-headed arrow in the inset of Fig.\,3(b) is attributed to the monopole contribution.  The total specific heat $C\approx 1$\,J/Yb-mol\,K at 0.6\,K (the inset of Fig.\,2) and $v$, which is roughly determined by $v\sim aJ_{z\pm}/2\pi \hbar \sim15$\,m/s, where  $a(=0.43$\,nm) is the distance between neighboring tetrahedra,
 yield $\ell \sim 100$\,nm, or equivalently the scattering time $\tau\sim 2.5$\,ns.  We stress that this long $\ell$ is still underestimated, since the total specific heat and the initial reduction of thermal conductivity give only an overestimate and underestimate, respectively, for the monopole contribution.  This indicates that the excitations are mobile to a very long distance, $\ell > 250a$, without being scattered.  We stress that $\ell$ is much longer than the inter-monopole distance, which is estimated to be at most $5a$, assuming monopole density of 1\% of total number of tetrahedra. This corresponds to a very large coherent volume including more than $\sim 10^7$ tetrahedra, demonstrating highly mobile transport of this long-lived particle, whose effective mass is as heavy as $\sim 2000$ times the bare electron mass \cite{Pan-15}. This extremely small scattering rate may be due to the quantum feature,  which prohibits the simple monopole-antimonopole pair annihilation that violates energy conservation. 
%The thermal conductivity of Yb$_2$Ti$_2$O$_7$ in the spin liquid regime is essentially different from that of classical spin ice system Dy$_2$Ti$_2$O$_7$, in which $\kappa(H)$ always shows monotonic field dependence.  In Dy$_2$Ti$_2$O$_7$ $\kappa(H)$ is closely related to  the magnetization curve at low field, while in the regime (i) of Fig.\,3c where substantial contribution of the quantum monopoles to the thermal conductivity is observed, $\kappa(H)$ shows striking deviation from the $H/T$-scaling curve shown in Fig.\,3c.  Although extremely long mean free path of classical monopoles have been suggested in Dy$_2$Ti$_2$O$_7$,  it is highly unlikey that dispersionless classical monopoles shows such a ballistic transport. 

The present results indicate that the quantum fluctuations dramatically change the nature of the monopole excitations.  We note that the observed nearly gapless excitations are not relevant to the ``photon" excitations predicted by Ref.~\cite{Shannon-PRL12,Benton-PRB12,Gingras-RPP14,Savary-PRL12,Harmele-PRB04}. This is because the characteristic photon energy $E_{photon}\approx  J_{\perp}^3/J_{\parallel}^2\sim0.05$\,K is one order of magnitude smaller than the present temperature range, and hence the strongly temperature dependent $\alpha$ is incompatible with the photon excitations. The highly mobile heavy quantum monopoles in the spin liquid state is the most salient feature of the elementary excitations in  frustrated magnetic pyrochlore systems with strong quantum fluctuations. Nearly ballistic propagation phenomena of fractionalized magnetic excitations in  spin-liquid states have been reported in  spin-1/2 1D Heisenberg chain~\cite{Sologubenko-PRL00,Kudo-JPSJ01} and 2D triangular lattice with antiferromagnetic interactions~\cite{Yamashita-Science10}.  In the former elementary excitation is spinon which obeys semion statistics~\cite{Haldane-PRL91} and in the latter excitation has been discussed in terms of spinon which obeys fermionic statistics~\cite{aho0,aho1,aho2,aho3,aho4,Watanabe-NatCommun12}. In the present 3D system elementary excitation in the spin liquid state is quantum monopole, which is another fractionalized spinon. The residual $\kappa /T\mid_{T\rightarrow 0\,{\rm K}}$, which is distinctly present in the 2D case \cite{Yamashita-Science10,Watanabe-NatCommun12}, is absent in Yb$_2$Ti$_2$O$_7$ (Fig.\,2(c)), implying that this 3D spinon is unlikely to be fermionic. In fact, bosonic spinon has been presumed theoretically in 3D pyrochlore lattice~\cite{Sentil}. In 1D Heisenberg system, the mean free path is infinite at nonzero temperature due to the integrability of the Hamiltonian. The highly mobile fermionic spinons in 2D and bosonic quantum monopoles in 3D may be a key feature of the elementary excitations in highly frustrated quantum magnets and its origin is an open question. 

We thank  L.\,Balents, K. Behnia, S.\,Fujimoto, H.\,Kawamura, S.\,Onoda, and K.\,Totsuka  for useful discussions. Financial support for this work was provided by Grants-in-Aid for Scientific Research from the Japan Society for the Promotion of Science (JSPS).
\\

\noindent



%%%%%%%%%% Merge with supplemental materials %%%%%%%%%%
\pagebreak
\begin{center}
\textbf{\large Supplemental Materials for "Thermal conductivity of quantum magnetic monopoles in the frustrated pyrochlore Yb$_2$Ti$_2$O$_7$"}
\end{center}
%%%%%%%%%% Merge with supplemental materials %%%%%%%%%%
%%%%%%%%%% Prefix a "S" to all equations, figures, tables and reset the counter %%%%%%%%%%
\setcounter{equation}{0}
\setcounter{figure}{0}
\setcounter{table}{0}
\setcounter{page}{1}
\makeatletter
\renewcommand{\theequation}{S\arabic{equation}}
\renewcommand{\thefigure}{S\arabic{figure}}
\renewcommand{\bibnumfmt}[1]{[S#1]}
\renewcommand{\citenumfont}[1]{S#1}
%%%%%%%%%% Prefix a "S" to all equations, figures, tables and reset the counter %%%%%%%%%%

\section{Additional phonon scatterings in D\MakeLowercase{y}$_2$T\MakeLowercase{i}$_2$O$_7$}
$\kappa/T$ of Y$_2$Ti$_2$O$_7$, Dy$_2$Ti$_2$O$_7$ at zero field and Yb$_2$Ti$_2$O$_7$ at $H$=12\,T are compared in Fig.S1. $\kappa/T$ of Dy$_2$Ti$_2$O$_7$ at high temperature is strongly suppressed from Y$_2$Ti$_2$O$_7$. It has been reported that $\kappa/T$ of Dy$_2$Ti$_2$O$_7$ decreases monotonically with $H$, indicating even smaller phonon $\kappa/T$~\cite{S1}. This implies the presence of unknown phonon scatterings, which are not likely caused by crystal field excitations because of the large energy gap of first excited state, $\sim$380\,K~\cite{S2}. As discussed in the main text, $\kappa/T$ of Yb$_2$Ti$_2$O$_7$ at $H$=12\,T represents the purely phononic $\kappa/T$ at low temperatures, where $k_{\rm B}T<<\mu_{\rm B}H$. $\kappa/T$ at 2\,K is much larger than Dy$_2$Ti$_2$O$_7$ and close to Y$_2$Ti$_2$O$_7$, suggesting the absence of such unknown scatterings. 

\begin{figure}[h]
\includegraphics[width=\linewidth,keepaspectratio]{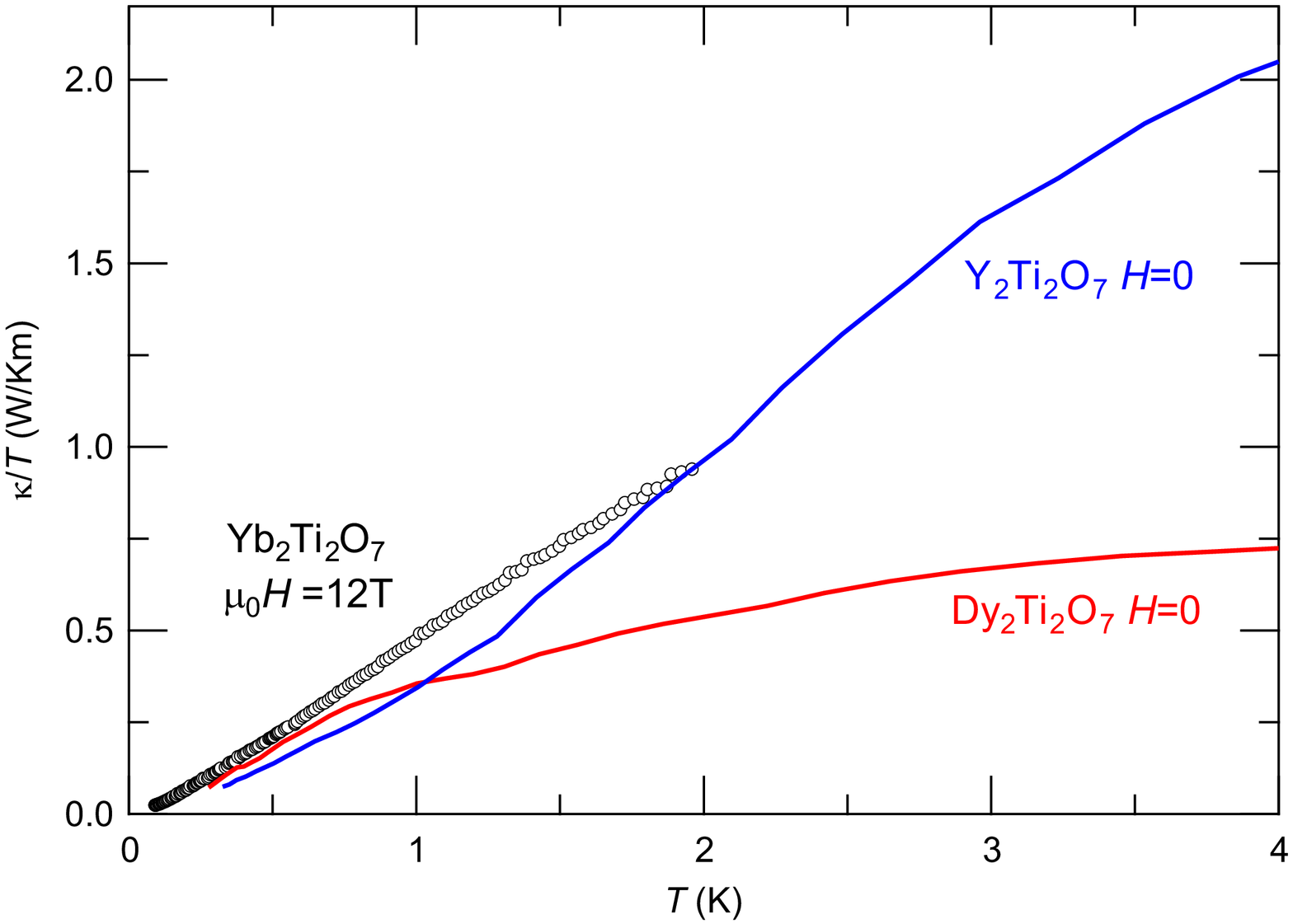}
\caption{(color online). $\kappa/T$ of Y$_2$Ti$_2$O$_7$, Dy$_2$Ti$_2$O$_7$ at zero field and Yb$_2$Ti$_2$O$_7$ at $H$=12\,T applied parallel to [1,1,1] with the same heat current direction along [1,-1,0]. Data of Y$_2$Ti$_2$O$_7$, Dy$_2$Ti$_2$O$_7$ are taken from Ref.~\cite{S1}.  }
\end{figure}

\section{Disappearance of initial reduction of $\bm{\kappa(H)/T}$ with $\bm{H}$ below $\bm{T_{\rm C}}$=0.19\,K}

\begin{figure}[t]
\includegraphics[width=\linewidth,keepaspectratio]{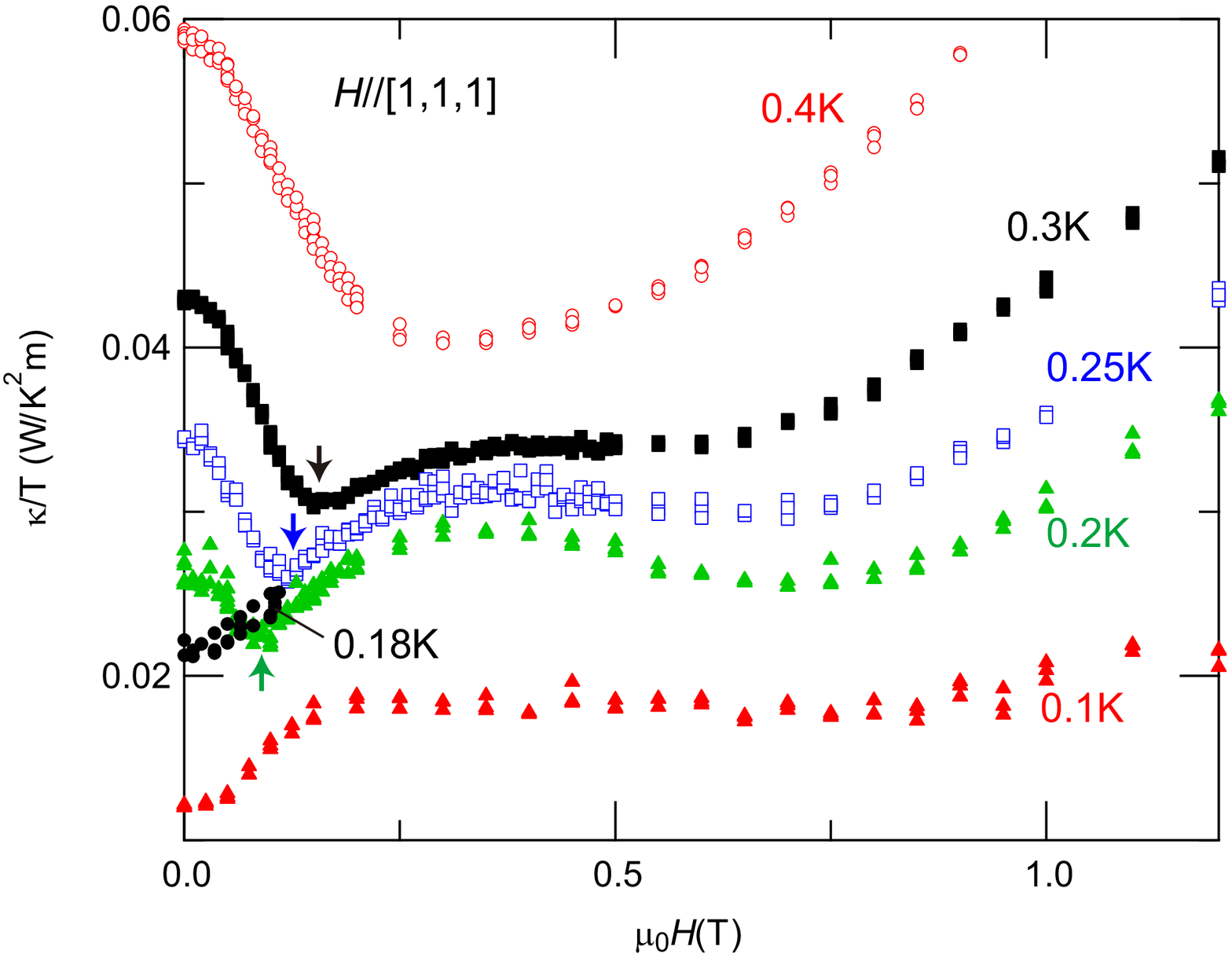}
\caption{(color online). Field dependence of $\kappa/T$ of Yb$_2$Ti$_2$O$_7$ for $H\parallel$[1,1,1] with the heat current along [1,-1,0]. Arrows indicate the field-induced transition from paramagnetic to ferromagnetic states.}
\end{figure}

\begin{figure}[h*]
\includegraphics[width=\linewidth,keepaspectratio]{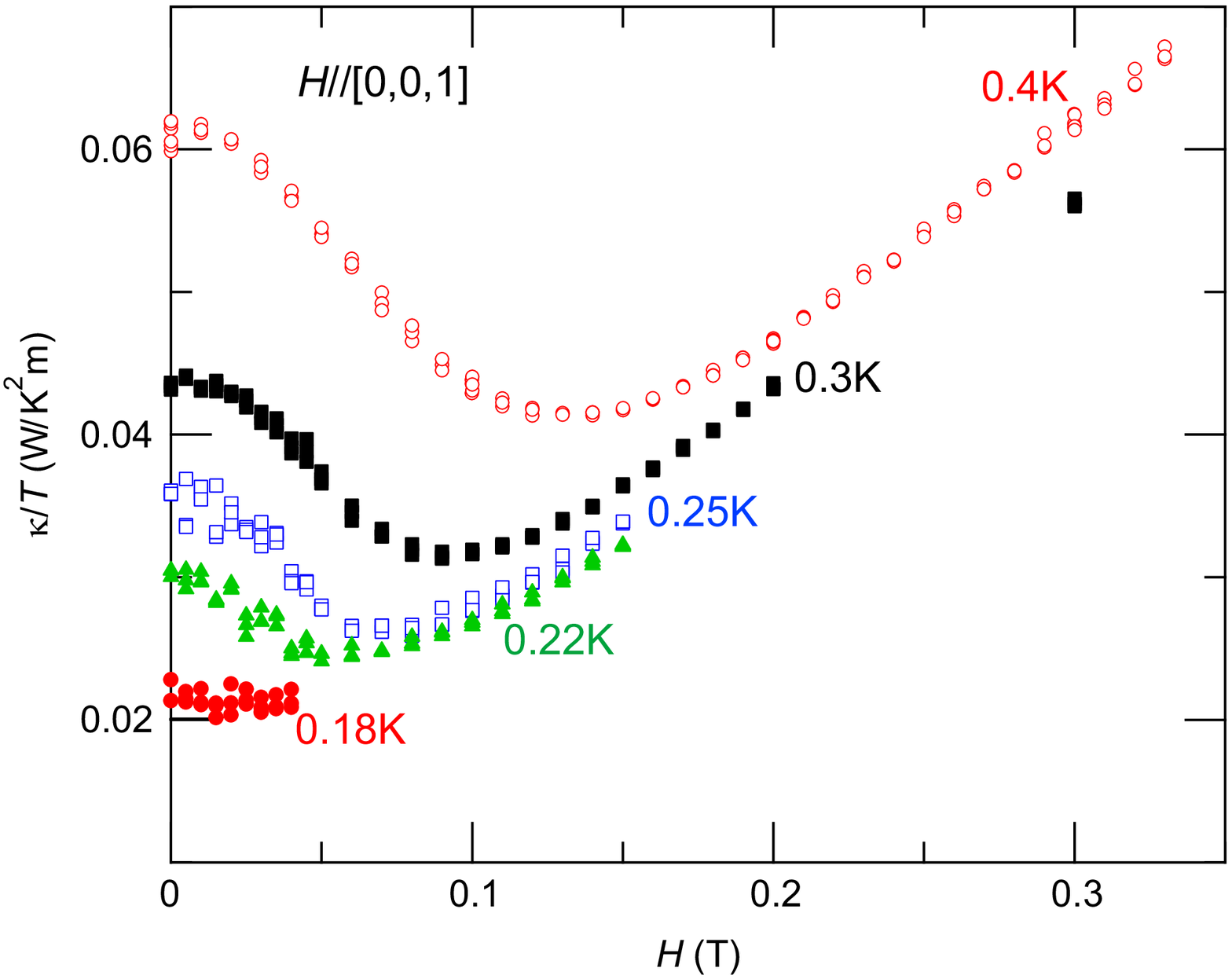}
\caption{(color online). Field dependence of $\kappa/T$ of Yb$_2$Ti$_2$O$_7$ for $H\parallel$[0,0,1] with the heat current along [1,-1,0] }
\end{figure}

In a temperature range of 0.2\,K$\leq T\leq$0.3\,K, a clear kink appears in $\kappa(H)/T$ for $H\parallel$[1,1,1] at the  transition between spin liquid and ferromagnetic states, indicated by arrows in Fig.\,S2. The kink is shifted to lower field with decreasing temperature and vanishes below $T_{\rm C}$ in the ferromagnetic state. The position of the kink agrees with the field dependence of $T_{\rm C}(H)$ determined by specific heat measurements, which increases with $H$ at low field region~\cite{S3}. 

As discussed in the main text, the initial reduction of $\kappa(H)/T$ with $H$ observed above $T_{\rm C}$ indicates thermal conduction of magnetic quantum monopoles. The initial reduction in the temperature range 0.2\,K$\leq T\leq$0.3\,K is interrupted by the field-induced ferromagnetic ordering and $\kappa(H)/T$ shows a characteristic enhancement with the field in the ferromagnetic state. The enhancement is understood by the suppression of elastic scattering of phonon due to the ordering of magnetic moments. As the kink disappears in the ferromagnetic state, the initial reduction, which is the signature of monopole heat conduction, also disappears, in consistent with the suppression of spin-ice correlations below $T_{\rm C}$ reported by neutron scattering experiments~\cite{S4}. The initial reduction disappears below $T_{\rm C}$ also for $H\parallel$[0,0,1]　(Fig.\,S3).

\renewcommand{\figurename}{TABLE}
\setcounter{figure}{0}
\begin{figure}[t]
\includegraphics[width=\linewidth,keepaspectratio]{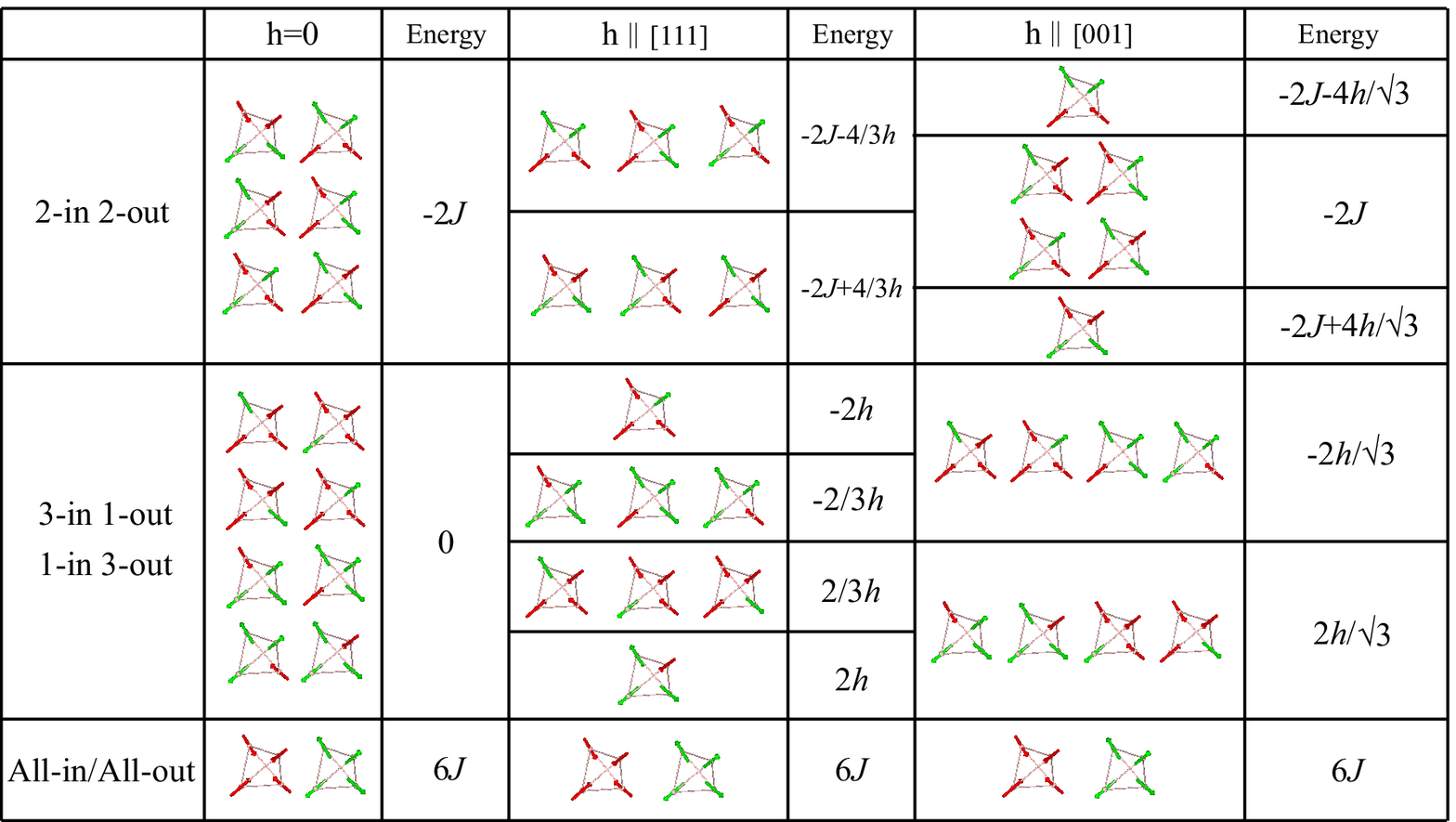}
\caption{(color online). The energy of spin configurations in a single tetrahedron}
\end{figure}
\setcounter{figure}{3}
\renewcommand{\figurename}{FIG.}

\section{Initial reduction of classical monopole density with magnetic field }

By calculating the classical monopole density ($\rho_{31}$, 3-in-1-out and 1-in-3-out configurations) in magnetic field, we show that $\rho_{31}$ decreases with $H^2$ at zero-field limit, regardless of the field direction. 

Hamiltonian of a nearest-neighbor spin ice model is written as
\begin{equation}
H=J\sum_{<i,j>}\sigma_i^z\sigma_j^z-g\mu_B\bm{H}\cdot\sum_j\bm{S}_j \nonumber
\end{equation}

where $J$, which corresponds to $J_{\parallel}$ in the main text, is the nearest neighbor Ising interaction, the spin $\bm{S}_j $ is an Ising spin: $\bm{S}_j =\sigma_j^z\bm{d}_j$, $\sigma_j^z=\pm 1$, with the anisotropy axes, $\bm{d}_0=[1,1,1]/\sqrt{3}$, $\bm{d}_1=[1,-1,-1]/\sqrt{3}$, $\bm{d}_2=[-1,1,-1]/\sqrt{3}$, $\bm{d}_3=[-1,-1,1]/\sqrt{3}$. With $h=g\mu_BH$, the energy of all the spin configurations are shown in Table S1. 

For $H\parallel$[1,1,1], $\rho_{31}$ is derived as,
\begin{eqnarray}
\rho_{31}=\frac{8\cosh^3⁡(2h_J/3t)}{N_{[1,1,1]}} \nonumber
\end{eqnarray}
\begin{eqnarray}
&N_{[1,1,1]}=6\exp⁡(2/t)\cosh⁡(4h_J/3t)\nonumber\\
&\hspace{2.5cm}+8\cosh^3(2h_J/3t)+2\exp⁡(-6/t) \nonumber
\end{eqnarray}
Here, $h_J=h/J$ and $t=k_BT/$$J$ are normalized field and temperature, respectively. 

\begin{figure}[t]
\includegraphics[width=\linewidth,keepaspectratio]{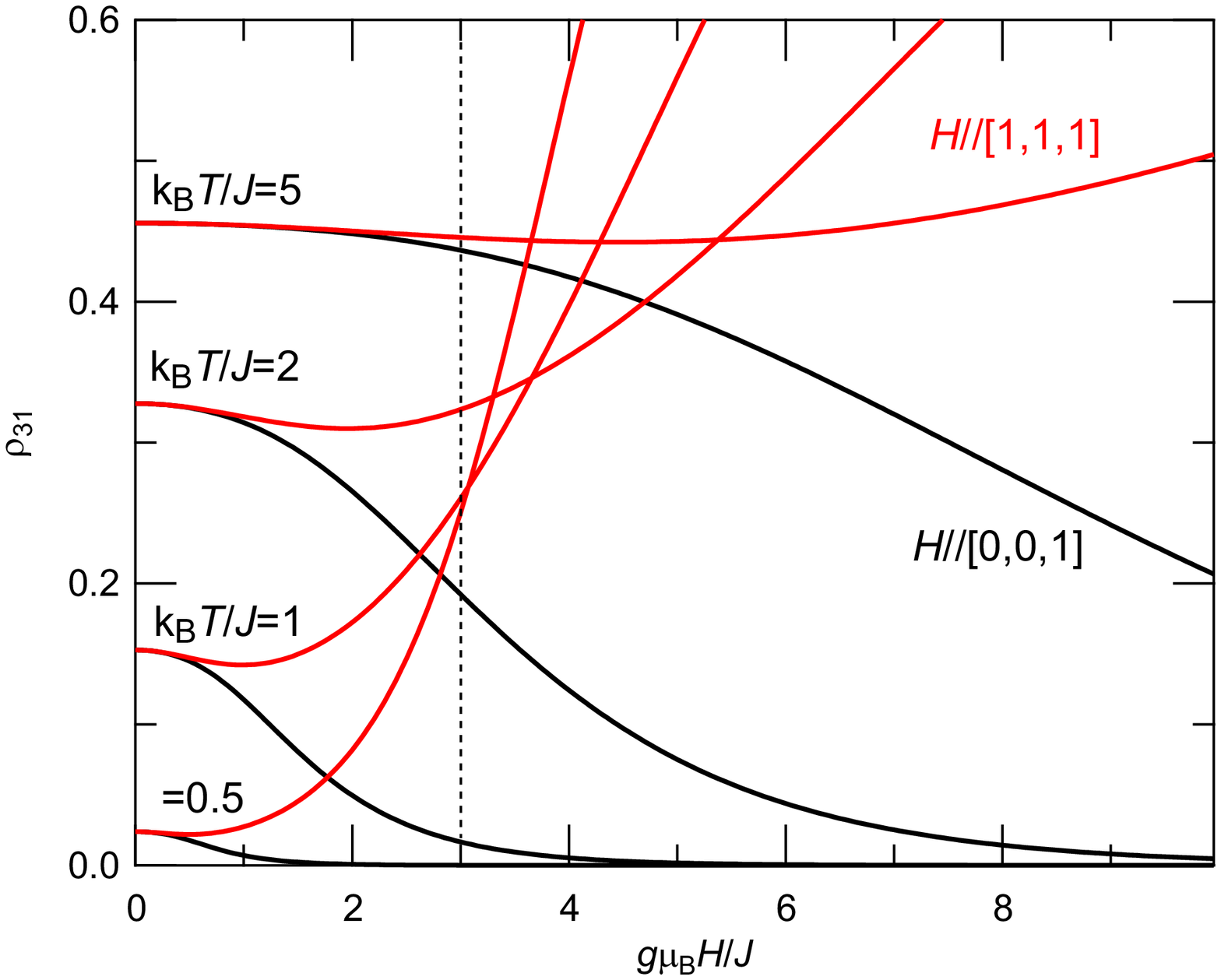}
\caption{(color online). Classical monopole density  $\rho_{31}$ is plotted against normalized magnetic field $g\mu_BH/J$ for $H\parallel$[1,1,1] and [0,0,1]. Dotted vertical line indicates the field $g\mu_BH/J$=3, at which level crossing of 3-in-1-out and 2-in-2-out occurs when $H$ is applied along [1,1,1].}
\end{figure}
\begin{figure}[h*]
\includegraphics[width=\linewidth,keepaspectratio]{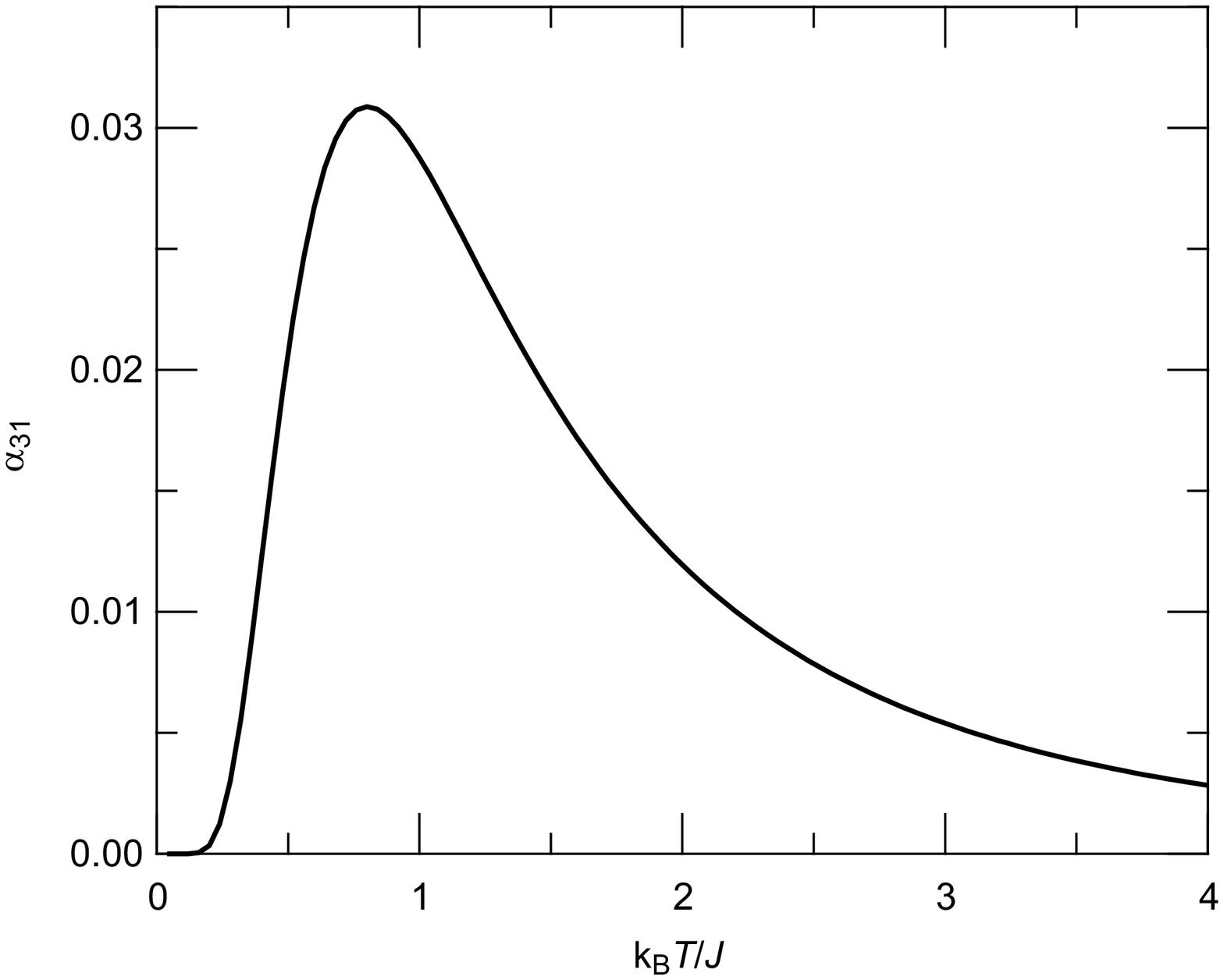}
\caption{(color online). $h_J^2$ coefficient of $\rho_{31}$, $\alpha_{31}$, is plotted against normalized temperature, $k_BT/J$.}
\end{figure}  

For $H\parallel$[0,0,1],
\begin{eqnarray}
\rho_{31}=\frac{8\cosh⁡(2h_J/\sqrt{3}t)}{N_{[0,0,1]}}\nonumber
\end{eqnarray}
\begin{eqnarray}
&N_{[0,0,1]}=2\exp⁡(2/t)[2+\cosh⁡(4h_J/\sqrt{3}t)]\nonumber\\
&\hspace{1.8cm}+8\cosh⁡(2h_J/\sqrt{3}t)+2\exp⁡(-6/t) \nonumber
\end{eqnarray}

The resulting field dependencies at different temperatures are plotted in Fig.S4. For $H\parallel$[0,0,1], the energy of one 2-in-2-out configuration decreases the most by Zeeman effect, leading to  monotonic increase of 2-in-2-out density. As a result, $\rho_{31}$ decreases monotonically. For $H\parallel$[1,1,1], the energy of one 3-in-1-out configuration decreases the most and crosses with the lowest energy of 2-in-2-out configuration at $g\mu_BH/J=$3. As this crossing occurs, $\rho_{31}$ rapidly increases with $H$. On the other hand, in low field region, $\rho_{31}$ decreases with field isotropically. We verify this by expanding $\rho_{31}$ with $h_J$ around $h_J$=0. For both the two field directions, $h_J$-linear term vanishes and $h_J^2$ term is identical. The isotropic field dependence of $\rho_{31}$ at zero-field limit is then,
\begin{eqnarray}
\rho_{31}(h_J)=\rho_{31}(0)-\alpha_{31}h_J^2…\nonumber
\end{eqnarray}
\begin{eqnarray}
\alpha_{31}=\frac{8(\exp⁡(2/t)-\exp⁡(-6/t))}{3t^2(\exp⁡(-6/t)+3\exp⁡(2/t)+4)^2}  \nonumber
\end{eqnarray}

The $h_J^2$ coefficient $\alpha_{31}$ is plotted against $t=k_BT/$$J$ in Fig.S5. It exhibits a maximum at $k_BT_{max,⁡             \alpha_{31}}/J$=0.8. The relation between $T_{max,\alpha_{31}}$ and the monopole excitation energy $\Delta_{31}$ is then $\Delta_{31}$=2$J$=2.5$k_BT_{max,⁡         \alpha_{31}}$. In the main text, the initial $H^2$ decrease of $\kappa/T$ is ascribed to decreasing number of monopoles and the $H^2$ coefficient $\alpha$ in the field dependence of $\kappa/T$ exhibits a maximum at $T_{max}$=0.3-0.5\,K. If $T_{max}$ is related to the monopole gap energy, it corresponds to 0.75-1.25K of monopole excitation energy, which is strongly suppressed from the classical one, 2$J_{\parallel}$=4\,K~\cite{S5}. It should be noted, however, that this estimation is of purely classical monopole.


\begin{thebibliography}{99}

\bibitem{Bramwell-Science01}
S. T. Bramwell, and M. J. P. Gingras, Science {\bf 294}, 1495 (2001).

\bibitem{Ramirez-Nature99}
A. P. Ramirez, A. Hayashi, R. J. Cava, R. Siddharthan, and B. S. Shastry, Nature {\bf 399}, 333 (1999).

\bibitem{Bramwell-PRL01}
S. T. Bramwell, M. J. Harris, B. C. den Hertog, M. J. P. Gingras, J. S. Gardner, D. F. McMorrow, A. R. Wildes, A. L. Cornelius, J. D. M. Champion, R. G. Melko, and T. Fennell, Phys. Rev. Lett. {\bf 87}, 047205 (2001).

\bibitem{Fennell-Science09}
T. Fennell, P. P. Deen, A. R. Wildes, K. Schmalzl, D. Prabhakaran, A. T. Boothroyd, R. J. Aldus, D. F. McMorrow, and S. T. Bramwell, Science {\bf 326}, 415 (2009).

\bibitem{Castelnovo-Nature08}
C. Castelnovo, R. Moessner, and S. Sondhi, Nature {\bf 451}, 42 (2008).

\bibitem{Morris-Science09}
D. J. P. Morris, D. A. Tennant, S. A. Grigera, B. Klemke, C. Castelnovo, R. Moessner, C. Czternasty, M. Meissner, K. C. Rule, J.-U. Hoffmann, K. Kiefer, S. Gerischer, D. Slobinsky and R.S. Perry, Science {\bf 326}, 411 (2009).

\bibitem{Shannon-PRL12}
N. Shannon, O. Sikora, F. Pollmann, K. Penc, and P. Fulde, Phys. Rev. Lett. {\bf 108}, 067204 (2012).

\bibitem{Benton-PRB12}
O. Benton, O. Sikora, and N. Shannon, Phys. Rev. B {\bf 86}, 075154 (2012).

\bibitem{Gingras-RPP14}
M. J. P. Gingras, and P. A. McClarty, Rep. Prog. Phys. {\bf 77}, 056501 (2014).

\bibitem{Savary-PRL12}
L. Savary, and L. Balents, Phys. Rev. Lett. {\bf 108}, 037202 (2012).

\bibitem{Harmele-PRB04}
M. Hermele, and M. P. A. Fisher, L. Balents, Phys. Rev. B {\bf 69}, 064404 (2004).

\bibitem{Ross-PRX11} K. A. Ross, L. Savary, B. D. Gaulin, and L. Balents, Phys. Rev. X {\bf 1}, 021002 (2011).

\bibitem{Applegate-PRL12} R. Applegate, N. R. Hayre, and R. R. P. Singh, T. Lin, A. G. R. Day, and M. J. P. Gingras, Phys. Rev. Lett. {\bf 109}, 097205 (2012).

\bibitem{Hirschberger-15} M. Hirschberger, J. W. Krizan, R. J. Cava, and N. P. Ong, arxiv:1502.02006 (2015).

\bibitem{Gardner-RMP10} J. S. Gardner, M. J. P. Gingras, and J. E. Greedan, Rev. Mod. Phys. {\bf 82}, 53 (2010). 

\bibitem{Hodges-JPhys01}
J. A. Hodges, P Bonville, A Forget, M Rams, K Kr\'{o}las, and G Dhalenne, J. Phys. Cond. Matt. {\bf 13}, 9301 (2001).

\bibitem{chang-NComm12}
L.-J. Chang S. Onoda, Y. Su, Y. -J. Kao, K. -D. Tsuei, Y. Yasui, K. Kakurai and M. R. Lees, Nat. Commun. {\bf 3}, 992 (2012).

\bibitem{Yasui-JPSJ03}
Y. Yasui, M.Soda, S. Iikubo, M. Ito, M. Sato, N. Hamaguch, T. Matsushita, N. Wada, T. Takeuchi, N. Aso, and K. Kakurai, J. Phys. Soc. Jpn. {\bf 72}, 3014 (2003).

\bibitem{Ross-PRB11}
K. A. Ross, L. R. Yaraskavitch, M. Laver, J. S. Gardner, J. A. Quilliam, S. Meng, J. B. Kycia, D. K. Singh, Th. Proffen, H. A. Dabkowska, and B. D. Gaulin, Phys. Rev. B {\bf 84}, 174442 (2011).

\bibitem{Hodges-PRL02}
J. A. Hodges, P. Bonville, A. Forget, A. Yaouanc, P. Dalmas de R\'{e}otier, G. Andr\'{e}, M. Rams, K. Kr\'{o}las, C. Ritter, P. C. M. Gubbens, C. T. Kaiser, P. J. C. King, and C. Baines, Phys. Rev. Lett. {\bf 88}, 077204 (2002).

\bibitem{supple}
See Supplemental Material for additional phonon scattering in Dy$_2$Ti$_2$O$_7$, disappearance of initial decrease of $\kappa$($H$) below $T_{\rm C}$ and theoretical field dependence of monopole density for classical spin ice.

\bibitem{Yaouanc-PRB11}
A. Yaouanc, P. Dalmas de R\'{e}otier, C. Marin, and V. Glazkov, Phys. Rev. B {\bf 84}, 172408
(2011).

\bibitem{Kolland-PRB12} 
G. Kolland, O. Breunig, M. Valldor,M. Hiertz, J. Frielingsdorf, and T. Lorenz, Phys. Rev. B {\bf 86}, 060402 (2012).

\bibitem{Kolland-PRB13} 
G. Kolland , M. Valldor, M. Hiertz, J. Frielingsdorf, and T. Lorenz, Phys. Rev. B {\bf 88}, 054406 (2013).

\bibitem{Udagawa}
M. Udagawa (unpublished)
%\bibitem{Ross-PRB12}
%K. A. Ross {\it et al}., Lightly stuffed pyrochlore structure of single-crystalline Yb$_2$Ti$_2$O$_7$ grown by the optical floating zone technique. {\it Phys. Rev. B} {\bf 86}, 174424 (2012).

\bibitem{Pan-15}
L. Pan, N. J. Laurita, Kate A. Ross, Edwin Kermarrec, Bruce D. Gaulin, and N. P. Armitage, arXiv:1501.05638 (2015).

\bibitem{Sologubenko-PRL00}
A. V. Sologubenko, K. Giann\'{o}, H. R. Ott, U. Ammerahl, and A. Revcolevschi, Phys. Rev. Lett. {\bf 84}, 2714 (2000).

\bibitem{Kudo-JPSJ01}
K. Kudo. S Ishikawa, T. Noji, T. Adachi, Y. Koike, K. Maki, S. Tsuji, and K. Kumagai, J. Phys. Soc. Jpn. {\bf 70}, 437 (2001).

\bibitem{Yamashita-Science10}
M. Yamashita, N. Nakata, Y. Senshu, M. Nagata, H. M. Yamamoto, R. Kato, T. Shibauchi, and Y. Matsuda, Science {\bf 328}, 1246 (2010).

\bibitem{Haldane-PRL91}
F. D. M. Haldane, Phys. Rev. Lett. {\bf 67}, 937 (1991).

\bibitem{aho0}
L. Balents, Nature {\bf 464}, 199-208 (2010).

\bibitem{aho1}
S.-S. Lee, and P. A. Lee, Phys. Rev. Lett. {\bf 95}, 036403 (2005).

\bibitem{aho2}
S.-S. Lee, P. A. Lee, and T. Senthil, Phys. Rev. Lett. {\bf 98}, 067006 (2007).

\bibitem{aho3}
M. S. Block, D. N. Sheng, O. I. Motrunich, and M. P. A. Fisher, Phys. Rev. Lett. {\bf 106}, 157202 (2011).

\bibitem{aho4}
M. Barkeshli, H. Yao, and S. A. Kivelson, Phys. Rev. B {\bf 87}, 140402 (2013).

\bibitem{Watanabe-NatCommun12}
D. Watanabe, M. Yamashita, S. Tonegawa,	Y. Oshima, H.M. Yamamoto,	R. Kato, I. Sheikin,	K. Behnia, T. Terashima,	S. Uji,	T. Shibauchi and Y. Matsuda, Nat. Commun. {\bf 3}, 1090 (2012).

\bibitem{Sentil}
T. Senthil, Discussion Meeting: Quantum entanglement in macroscopic matter.  (http://www.icts.res.in/lecture/details/1642/).

\end{thebibliography}

\begin{thebibliography}{11}

\bibitem{S1} 
G. Kolland , M. Valldor, M. Hiertz, J. Frielingsdorf, and T. Lorenz, Phys. Rev. B {\bf 88}, 054406 (2013).

\bibitem{S2} 
S. Rosenkranz, A. P. Ramirez, A. Hayashi, R. J. Cava, R. Siddharthan and B. S. Shastry, J. Appl. Phys. {\bf 87}, 5914 (2000)

\bibitem{S3} 
Y. Yasui et al., in preparation.

\bibitem{S4} 
L.-J. Chang S. Onoda, Y. Su, Y. -J. Kao, K. -D. Tsuei, Y. Yasui, K. Kakurai and M. R. Lees, Nat. Commun. {\bf 3}, 992 (2012).

\bibitem{S5} 
K. A. Ross, L. Savary, B. D. Gaulin, and L. Balents, Phys. Rev. X {\bf 1}, 021002 (2011).

\end{thebibliography}
\end{document}